
%
%
%
\def\figflag{I}                                                           %
%
%

\input harvmac.tex
%
%
\def\Fig#1{Fig.~\the\figno\xdef#1{Fig.~\the\figno}\global\advance\figno
 by1}
\def\figI{I}
%
%
\newdimen\tempszb \newdimen\tempszc \newdimen\tempszd \newdimen\tempsze
\ifx\figflag\figI
\input epsf
%
\def\epsfsize#1#2{\expandafter\epsfxsize{
 \tempszb=#1 \tempszd=#2 \tempsze=\epsfxsize
     \tempszc=\tempszb \divide\tempszc\tempszd
     \tempsze=\epsfysize \multiply\tempsze\tempszc
     \multiply\tempszc\tempszd \advance\tempszb-\tempszc
     \tempszc=\epsfysize
     \loop \advance\tempszb\tempszb \divide\tempszc 2
     \ifnum\tempszc>0
        \ifnum\tempszb<\tempszd\else
           \advance\tempszb-\tempszd \advance\tempsze\tempszc \fi
     \repeat
\ifnum\tempsze>\hsize\global\epsfxsize=\hsize\global\epsfysize=0pt\else\fi}}
\epsfverbosetrue
\fi
%
%
%
%
\def\ifigure#1#2#3#4{
\midinsert
\vbox to #4truein{\ifx\figflag\figI
\vfil\centerline{\epsfysize=#4truein\epsfbox{#3}}\fi}
\narrower\narrower\noindent{\footnotefont
{\bf #1:}  #2\par}
\endinsert
}
%
%
\lref\haw {S.W. Hawking, Comm. Math. Phys. {\bf 43}  (1975) 199.}
\lref\pres {J. Preskill, Caltech preprint CALT-68-1819 hep-th/9209058}
\lref\wit {E. Witten, Phys. Rev. {\bf {D44}} (1991) 314.}
\lref\hv {H. Verlinde, ``Black Holes and Strings in Two Dimensions'',
 in the proceeding of the Sixth Marcel Grossman Meeting,
 World Scientific (1992).}
\lref\vv {E. Verlinde and H.Verlinde, Nucl. Phys. {\bf B 406} (1993) 43}
\lref\svv {K. Schoutens, E. Verlinde and H. Verlinde, Phys. Rev. {\bf D 48}
(1993) 2690.}
\lref\cghs {C. Callan, S. Giddings, J. Harvey, and A. Strominger,
 Phys. Rev. {\bf{D45}} (1992) 1005.}
\lref\rst {J.G. Russo, L. Susskind, L. Thorlacius, Phys. Rev. {\bf D47}
(1993) 533}
\lref\jhs {J. Harvey and A. Strominger, in {\it String Theory and
 Quantum Gravity}, Proceedings of 1992 Trieste Spring School,
 (World Scientific, 1993).}
\lref\str {A. Strominger, Phys. Rev. {\bf D 46} (1992) 4396.}
\lref\cj {D. Cangemi and R. Jackiw, MIT preprint, CTP 2165, hep-th/9302026
(February 1993)}
\lref\thooft{G. 't Hooft, Nucl. Phys. {\bf B335} (1990) 138.}
\lref\msw {G. Mandal, A Sengupta, and S. Wadia, Mod. Phys. Lett. {\bf A6}
(1991) 1685}
\lref\withole {E. Witten, Phys. Rev. {\bf{D44}} (1991) 314}
\lref\bical{A. Bilal and C. Callan, Nucl. Phys. {\bf B 394} (1993) 73.}
\lref\DDF{E. Del Giudice, P. Di Vecchia and S. Fubini, Ann. Phys. {\bf 70}
(1972) 378.}
\lref\comp{L. Susskind, L. Thorlacius and J. Uglum, Stanford preprint
SU-ITP-93-15.}
\lref\brower{R.C. Brower, Phys. Rev. {\bf D6} (1972) 1655.}
\lref\gn {S. Giddings and W. Nelson, Phys. Rev. {\bf D 46} (1992) 2486.}
\lref\hks {S. Hirano, Y. Kazama, Y. Satoh, preprint UT-Komaba 93-3}
\lref\birdav {See e.g. N. Birrell and P. Davies, {\it Quantum Fields in
Curved Space} (Cambridge, 1982) and references therein.
For more recent discussions of the moving mirror-black hole analogy see
R. Carlitz and S. Willey, Phys. Rev {\bf D 36} (1987) 2327, 2336, and
F. Wilczek, IAS-preprint HEP-93/12, hep-th/9302096 (February 1993).}

\def\X{{\rm \, x}}

\def\R{{\rm \, r}}
\def\T{{\rm \, t}}

\def\PP{{\rm p}}

\Title{\vbox{\baselineskip12pt\hbox{PUPT-1430}
    \hbox{October 1993}
    \hbox{hep-th/9311007}
    }}
{\vbox{\centerline{Dynamical Moving Mirrors}
	\vskip 2mm\centerline{ and Black Holes}}}

\vskip .35cm
\centerline{Tze-Dan Chung\footnote{$^\dagger$}
{\it chung@puhep1.princeton.edu}
 and Herman Verlinde}
\bigskip\centerline{\it Joseph Henry Laboratories}
\centerline{\it Princeton University}
\centerline{\it Princeton, NJ 08544}

\vskip 1.2cm
\rm
\noindent

\centerline{\bf Abstract}

A simple quantum mechanical model of $N$ free scalar fields
interacting with a dynamical moving mirror is formulated and
shown to be equivalent to two-dimensional dilaton gravity.
We derive the semi-classical dynamics of this system, by including
the back reaction due to the quantum radiation.
We develop a hamiltonian formalism that describes the time evolution
as seen by an asymptotic observer, and write
a scattering equation that relates the in-falling and out-going modes
at low energies. At higher incoming energy flux, however, the semi-classical
model appears to become unstable and the mirror seems to accelerate
forever along a trajectory that runs off to infinity. This instability
provides a useful paradigm for black hole formation and introduces
an analogous information paradox. Finally, we indicate a possible
mechanism that may restore the stability of the system at the quantum
level without destroying quantum coherence.

\Date {}

\newsec {Introduction}

Soon after the discovery of black hole emission effect by Hawking
\haw, it was realized that this effect was an example of a wider class
of phenomena in which particle creation occurs due to the observer
dependence of the vacuum. A particularly useful analogy, that is often
made, is with particle creation due to an accelerating mirror \birdav.
The formal
relation between the two systems becomes most transparent when
one considers the trajectory of the origin of the coordinate system
in the spherically reduced
black hole geometry. Mathematically, this point indeed behaves as a
reflecting mirror, and, due to the distortion of space-time near the
black hole, it will at late times seem to rapidly recede away from
the asymptotic observers with a constant
acceleration of $a = {1\over 8M}$. From this perspective, the black
hole emission effect arises due to the distortion of the incoming
vacuum after reflection of this accelerating mirror, producing
an apparently thermal outgoing state with temperature
$T_H = {1\over 8\pi M}$.
Moreover,  the out-going radiation will appear to describe a mixed state,
because signals that are sent in at late times will never seem to reach
the mirror point and thus never seem to be reflected back into
out-going signals.

While in general the moving mirror analogy is of course incomplete,
the correspondence becomes almost exact
in the context of the two-dimensional dilaton gravity models
of black holes. In these toy models
one typically considers massless matter fields, because
this leads to the technical simplification that their propagation
does not depend on the Weyl factor of the metric \cghs. However,
this tells us that the gravitational and matter fields interact
directly only via the reflection at the boundary point,
that plays the role of the origin $r=0$
in the analogy with the $s$-wave reduced Einstein theory.
All two-dimensional dilaton gravity models of this type \rst are therefore
physically equivalent to a model of free matter fields reflecting off
a dynamical moving mirror \vv.

In this paper we will further develop this reformulation of two-dimensional
dilaton gravity. We will describe both the classical and the semi-classical
dynamics of the matter-mirror system in the large $N$ limit, and in
particular derive the explicit form of the back reaction due to the
quantum radiation. We will find that there exists a range of
parameters for which there exists a well-defined scattering
equation relating the low energy incoming and out-going matter waves.
This subcritical $S$-matrix is energy preserving and
includes a regime in which most of the outgoing radiation looks thermal but
still describes a pure state.  At higher incoming energy flux, however,
black hole formation sets in and the semi-classical mirror trajectory
degenerates.
This super-critical case is of course the most interesting, because we
are then dealing with the black hole evaporation phenomenon and
confronted with the puzzle of information loss. In the first part of
this paper, however, we will restrict our attention to the low
energy regime,
with the hope that a good understanding of this regime will
teach us something about how to extend the model to the super-critical
situation.

In particular we would like to gain insight into how one can
ensure {\it in a natural way} that the total energy carried by the emitted
radiation is equal to the total energy of the incoming matter. As a first
step in this direction, we will derive a hamiltonian that generates
the classical time evolution of the mirror-matter system as seen
by the asymptotic observer. We will find, however, that this hamiltonian
is unbounded from below and that the resulting instability leads to
black hole formation at high energies. In our model, this situation
is described by a forever accelerating mirror trajectory, which also
seems to radiate forever. In a concluding section we then explain why
we believe that this instability will be cured once all interactions
between the in-falling and out-going matter are taken into account.
In particular we point out that this interaction results in a
non-local commutator algebra between the $in$ and $out$-fields, and
this leads to some important modifications of the standard
semi-classical description.

\newsec{Back Reaction on a Moving Mirror.}

Before we turn to the study of two-dimensional dilaton gravity, let us
first discuss in general the effect of the back reaction due to
the emitted radiation in a moving mirror model. This discussion will
be a useful preparation for the coming sections and at the same time
it will summarize some of the conclusions of the more detailed
subsequent analysis.

Let us consider a two-dimensional
system of a relativistic particle of mass $m$ that acts as
a reflecting mirror for a collection of $N$ massless
matter fields $f_i$.
We can specify the motion of the mirror particle by the parametrized
world line trajectory $(\X^+(\tau), \X^-(\tau))$, and we will
assume that the matter fields are restricted to live to the right
of this world line. In other words, the mirror particle represents
the boundary of space, which has therefore only one asymptotic region.
For simplicity, let us choose the direct reflection condition that the
matter fields $f_i$ vanish along the boundary trajectory
\eqn\reff{f_i(\X^+(\tau)), \X^-(\tau)) = 0}
for all $\tau$. For a given
mirror trajectory, this condition relates
the left-moving in-modes and right-moving out-modes via the diffeomorphism
\eqn\scat{f_i^{out}(x^-) = f_i^{in}(F(x^-)),}
with
\eqn\diffe{\X^+(\tau) = F(\X^-(\tau)).}
It will be convenient to fix the freedom in the parameter $\tau$,
by identifying it with the proper time along the world line
\eqn\fix{ {d \X \over d \tau}^+{d \X \over d \tau}^- = 1.}
The equation of motion of the mirror particle then takes the general
form
\eqn\eomp{{m\over 2} {d^2 \X^\pm \over d\tau^2} = {\cal F}_\mp}
where ${\cal F}_\mp$ defines the relativistic force. From \fix \ we know
that force ${\cal F}$ is always perpendicular to the world line trajectory
\eqn\work{{\cal F}_+ {d \X}^+ + {\cal F}_-{d \X}^- = 0.}
with $d\X^\pm = {d\X\over d\tau}^\pm d\tau$.

In general there can be many external forces that
contribute in \eomp. These external forces could e.g. be used to generate
a non-zero acceleration of the mirror particle even in the absence of
(classical) matter waves. As a concrete example, that will enable us
to make a direct correspondence with two-dimensional dilaton gravity, let
us assume the mirror particle has a charge (set equal to 1) and moves in
a constant electric field $E$
\eqn\forc{{\cal F}^e_\pm = \pm E {d\X\over d\tau}^\mp.}
The vacuum mirror trajectories in the presence of this constant Lorentz force
are of the form
\eqn\hypo{(\X^+ - c^+)(\X^--c^-) = {m^2 \over 4 E^2}}
with $c^\pm$ integration constants. The sign of ${E\over m}$ determines
which branch of this hyperbola is picked out.

In non-vacuum situations, there will be an additional force
on the boundary particle due to the interaction with the massless scalar
fields. Each time an $f$-particle bounces off the mirror it will give
off some of its momentum, resulting in a variable force equal to the rate
of change in the momentum in the matter fields
\eqn\force{{\cal F}^m_\pm = \pm  T_{\pm\pm}
{d \X \over d\tau}^\pm.}
Here
\eqn\stress{T_{\pm\pm} = \sum_i \half (\partial_\pm f_i)^2}
denotes the traceless matter stress-energy tensor. The orthogonality
condition \work \ on the force is equivalent to the conservation
of stress-energy
\eqn\refl{T_{--} ({d\X}^-)^2
= T_{++}({d\X}^+)^2}
This conservation equation is classically consistent
with the reflection condition \reff.
The resulting equation of motion \eomp\ in the presence of the combined
force ${\cal F}^m_\pm + {\cal F}^e_\pm$
becomes, after integrating once with respect to $\tau$
\eqn\momc{ {m\over 2}
{d\X\over d\tau}^\pm  \pm E \X^\pm +  P_\mp(\X^\mp) = {\rm constant}}
with
\eqn\ppdef{P_\pm(\X^\pm) = \pm \int\limits_{{\X^\pm}}^{\pm\infty} \!\!
dx^\pm\, T_{\pm\pm}.}
These equations express the classical
conservation of total energy and momentum.

Now we would like to incorporate into this simple model the quantum
mechanical effect of particle creation by a moving mirror.
It is well known that a
diffeomorphism of the form \scat\ will in general mix the positive and
negative energy modes of the $f_i$-fields, and that, as a result,
an incoming vacuum state will be reflected to an out-going state with
a non-zero flux of particles.
To account for the energy of these particles, the classical
reflection equation \refl\ for the stress-energy tensor will receive a
quantum correction. For massless fields in two-dimensions, this
correction is given by the conformal anomaly
\eqn\reflq{T_{--} ({d\X}^-)^2
= (T_{++} + {\kappa} \{ \X^-,\X^+\} )({d\X}^+)^2}
where $\kappa = {N\over 24}$ and
$\{ \ , \ \}$ denotes the schwartzian derivative
\eqn\schw{ \{ f, x \} =   {f^{\prime\prime\prime}\over
f^\prime}  -{3\over 2} {(f^{\prime\prime})^2\over
(f^\prime)^2}}
with $f^\prime = {df\over dx}$. This quantum reflection equation shows
that if we start with vacuum with $T_{++}\!=\! 0$,
then after reflection there
is a non-zero out-going energy flux $T_{--} \! =\!  \kappa \{\X^+,\X^-\}$.

This particle creation phenomenon will result in a modification of
the equation of motion for the mirror, to account for the back reaction
due to the quantum radiation. Indeed, with the new
reflection equation \reflq, the classical expression \force\ for
the force is no longer consistent with the transversality condition \work.
To correct for this, we should rewrite the new
equation \reflq\ in a similar form as the old one \refl, because only then
we can consistently interpret each side of the equation as the quantum
corrected force to be used in \eomp. Unfortunately, however, this procedure
is not entirely unique, and thus there seems to be more than one way
to include a semi-classical correction to the force equation.  Presumably,
this ambiguity is related to the freedom one has in writing the one-loop
counterterm that compensates for the conformal anomaly (see section 5
for a more detailed discussion of this point).

The most convenient choice, that
leads to the simplest equations of motion for the boundary, is to
rewrite \reflq\ as follows
\eqn\reflq{(T_{--} + {\kappa\over 2} \partial_-^2 \log \partial_- \X^+)
({d\X}^-)^2
= (T_{++} + {\kappa\over 2} \partial_+^2 \log \partial_+\X^- )
({d\X^+})^2}
Thus, comparing with \work, this choice leads to the following
expression for the quantum corrected force
\eqn\forceq{\eqalign{{\cal F}_\pm&=\pm(T_{\pm\pm} + {\kappa\over 2}
\partial_\pm^2\log( \partial_\pm\X^\mp)){d \X\over d\tau}^\pm\cr
&={d\over d\tau}(P_\pm(\X^\pm)\pm{\kappa}{d^2\X\over d\tau^2}^\pm)}}
After inserting this result into \eomp\ we obtain a quantum corrected
equation of motion. The fact that it becomes a third order differential
equation is typical for systems where the emitted radiation depends
and reacts back on the motion of the source.  As before, however, we
can integrate the equation of motion once with respect to the
proper time $\tau$ and obtain second order
differential equations
\eqn\momc{{m\over 2}{d\X\over d\tau}^\pm \pm
{\kappa}{d^2\X\over d\tau^2}^\pm
\pm E\X^\pm +  P_\mp(\X^\mp)={\rm constant}}
where we recall that $\tau$ is defined as the proper time along the boundary.
The above equations can in principle be used to compute an explicit
scattering equation describing the reflection of $f$ waves off the
dynamical mirror.

In the following sections we will examine two-dimensional dilaton gravity
and we will arrive at an essentially identical set of equations of motion
for the boundary point. An unusual feature, however, will be that
the mirror particle in that case has a negative rest mass.
We will discuss the properties of the above equations in more detail
in section 6.

\newsec{Dilaton Gravity as a Moving Mirror Model}

In this section we describe the reformulation of two-dimensional dilaton
gravity as a simple two-dimensional model of free scalar fields
interacting with a dynamical moving mirror. Our discussion here
will be purely classical. The semi-classical corrections due to
the conformal anomaly will be discussed in section 5 and 6.

\subsec {Classical dilaton gravity}

Two-dimensional
dilaton gravity coupled to $N$ massless scalar fields is described by
the action
\eqn\scg{
S_0 = {1\over 2\pi}\int d^2x\sqrt{-g} [e^{-2\phi}(R + 4(\nabla\phi)^2+
4\lambda^2) + {1\over 2}\sum_{i=1}^N(\nabla f_i)^2 ]}
This action shares many features with the
$s$-wave reduction of 3+1-dimensional
gravity  and has been extensively studied as a toy model of two-dimensional
quantum gravity which contains black hole solutions.
The classical equations of motion of this action can be solved explicitly
due to the property that the rescaled metric
\eqn\resc{\hat{g}_{ab} = e^{2\phi}g_{ab}}
is flat everywhere. To write the general solution,  it is thus natural
to introduce coordinates $x^\pm$ such that
\eqn\met{ds^2 = e^{2\rho(x^+,x^-)} dx^+ dx^-}
\eqn\roisvi{\rho(x^+,x^-)= \phi(x^+,x^-).}
In these coordinates the remaining equations of motion
\eqn\eom{
\partial_+\partial_- e^{-2\phi} = -\lambda^2}
\eqn\eomt{\partial^2_\pm e^{-2\phi} = T_{\pm\pm},}
where $T_{\pm\pm}$ 
is the traceless matter energy-momentum tensor,
are trivial to integrate to obtain the most general solution
\eqn\e{
e^{-2\phi}(x^+,x^-) =
 M -
\lambda^2 x^+ x^- - \int_{{}_{\! x^+}}^{{}^{\infty }}\!\!\!
dy^+\int_{{}_{\! y^+}}^{{}^{\infty}} \!\!\! dz^+ T_{++}
- \int^{{}^{ x^-}}_{{}_{\! -\infty}} \!\!\!
dy^-\int^{{}^{ y^-}}_{{}_{\!-\infty}} \!\!\! dz^- T_{--}}
If we put the $T_{\pm\pm}=0$ this reduces to the static black hole solution
of mass $M$. The terms involving $T_{\pm\pm}$
represent the classical back reaction of the metric
due to the incoming or outgoing matter.
The physical interpretation of this general solution as describing
gravitational
collapse and black hole formation in two-dimensions has been discussed
extensively elsewhere (for a review see \jhs).

Without matter present, the metric $ds^2$ reduces
to the two-dimensional Minkowski metric
\eqn\rind
{ds^2 = {dx^+ dx^- \over x^+x^-} }
with the two
light-cone coordinates $x^\pm$ defined on the half-lines
$\pm x^\pm>0$.
In this parametrization, the four asymptotic regions of the
Minkowski plane correspond to $x^\pm \rightarrow 0$ or $\pm \infty$.
In the following, however, we will consider only
the right regions as asymptotic regions, while the left regions
$x^\pm \rightarrow 0$ will be replaced by a dynamical boundary.
Namely, since the field $e^{\phi}$ is known to play the role of coupling
constant, it seems a natural procedure \rst \vv
to define a cutoff at strong coupling
by introducing a reflecting boundary located on a line
on which $e^{\phi}$ takes a certain large but constant value $e^{\phi_{cr}}$.
We can think of this dynamical boundary as describing the trajectory
of a single particle, parametrized by two coordinate
functions $\X^\pm(\tau)$ of an (arbitrary) time-variable $\tau$.
The condition that the dilaton is constant along the boundary
\eqn\bond{
\phi(\X^+(\tau),\X^-(\tau)) = \phi_{cr}}
will imply a specific equation of motion for its trajectory, provided
we supplement it with an appropriate reflection condition for the
matter fields. For most of the following discussion, we will keep the
specific value of $\phi_{cr}$ as a free parameter of the model.

If we set things up in this way, there will exist a low energy regime
in which this boundary remains time-like everywhere.
As long as we restrict ourselves to this regime,
we are allowed to impose the condition that the stress-energy
must reflect at the boundary. This guarantees that energy is conserved.
In this section we will work with the classical
reflection formula \refl. The quantum correction term will be included later.
There will of course also
be a high energy regime at which black hole formation sets in. In this case
part of the boundary is replaced by a space-like singularity. The critical
energy flux above which this happens depends on which critical value of
$\phi$ we choose in \bond, (see section 3.3). We will
comment on the super-critical case in the concluding section 7.

\subsec {Classical dynamics of the boundary}

The boundary
trajectory $\X^\pm(\tau)$ represents the only dilaton gravity degree of
freedom that couples directly to the matter fields. So in principle
one should be allowed to eliminate all other gravitational fields
from the action and derive a simple reparametrization invariant action
describing the dynamics of the boundary particle. Such a derivation,
however, would be slightly ambiguous, since it would depend on the
choice of boundary terms one could add to the original dilaton
gravity action. We will therefore follow a more direct route.

The boundary conditions \bond \ and \refl, together
with the explicit solution \e \ for the dilaton field, uniquely
determine the boundary equations of motion. We will now show that
these equations are the same as those derived from the following
very simple boundary action
\eqn\sbound{
S_b[\X] = m \int d \tau \sqrt{\partial_\tau {\X}^+ \partial_\tau
{\X}^-} - \lambda^2\int\! d\tau \,\X^+\partial_\tau {\X}^-
}
where the coupling of this boundary particle to matter is simply
described via the restriction that matter fields must live to the
right of the boundary. So the total action of the system is
\eqn\stot{ S = S_b[\X] + S_m[f, \X]}
with $S_b[\X]$ as above and
\eqn\sfree{
S_m[f,\X] = \int\limits_{{\pm\X^\pm < \pm x^\pm}}
\!\!\!\!\!\!\!\!\! d^2 x \, \partial_+ f \partial_-f}
where the integral over $x^\pm$ in the matter action is restricted
by the boundary as indicated.
The action \sbound\
is identical to that of a single particle moving in a constant
electric field $E=\lambda^2$. Note, however, that the coordinates
$x^\pm$ are not the usual Minkowski coordinates, for which the
metric takes the standard form
$ds^2 =  - dt^2 + dr^2$, but are (in the asymptotic region) related to
these via
\eqn\rt{r\pm t = \lambda^{-1}\log(\pm \lambda x^\pm).}

We have not yet put any restrictions on the sign of the
parameter $m$, and it in fact turns out that the relevant regime
corresponds to a particle of {\it negative} mass.
Namely, the free classical boundary trajectories that follow from
the action \sbound\
are hyperbolae of the form \hypo\ with $E = \lambda^2$ and
to select the correct branch requires that we take ${E\over m}<0$.
In the following we will therefore replace $m$ to $-m$, so that $m$
will continue to be a positive number, equal to minus the negative
mass of the mirror particle.
Further, we will be using a definite prescription
for fixing the integration constants $c^\pm$ in \hypo\
by imposing the asymptotic condition
that the past and future asymptotes of the boundary trajectory
are given by $x^+ = 0$ and $x^- = 0$, respectively.

The influence on the form of the boundary trajectory due to a
non-vanishing influx of energy is obtained by
computing the variation of the matter action with respect
to $\X^\pm$. One finds
\eqn\dels{\delta S_m = \int\! d\tau\,
\Bigl(T_{++} {\partial_\tau {\X}^+}\delta \X^+ +T_{--}
{\partial_\tau{\X}^-} \delta \X^-\Bigr).}
Integrating the resulting equation of motion $\delta S_b + \delta S_m = 0$
once with respect to $\tau$ gives
\eqn\bound{
-{m\over 2}\sqrt{\partial_\mp\X^\pm}
\pm \lambda^2 \X^\pm + P_\mp(\X^\mp) = 0.}
with $P_\pm$ as defined in \ppdef.
As in section 2, the above two equations express
the conservation of total $x^\pm$ momentum. Note, however, that
$P_\pm(\X^\pm)$ are not the usual total momentum when translated
back to $(r,t)$ coordinates.

The equations \bound\ combined imply that
\eqn\dif{
\partial_\tau \X^+ (\lambda^2 \X^- - P_+(\X^-)) + \partial_\tau
\X^- (\lambda^2 \X^+ + P_-(\X^+)) = 0.}
vThis is, as promised, the equation that the dilaton field,
as given via its classical solution \e, is constant along the boundary.
The value of this constant depends on the choice of the parameter $m$ via
\eqn\dilb{e^{-2\phi}(\X^+,\X^-) = {m^2\over 4\lambda^2 } .}

As a further comment we note that,
using the energy reflection equation, one can show that
also the following quantity
\eqn\fdm{
{\cal M}(\X^+,\X^-) =
M_+(\X^+) + M_-(\X^-)+{1\over \lambda^2}P_+(\X^+)P_-(\X^-),}
with
\eqn\mpdef{
M_\pm(\X^\pm)  =\int\limits_{\X^\pm}^{\pm\infty}\!\! dx^\pm
x^\pm T_{\pm\pm}.}
is conserved. This quantity ${\cal M}$ reduces in the far past and future
to the total energy as measured by an asymptotic observer. The result that
${\cal M}$ is constant can therefore be interpreted as the statement
of energy conservation.

\ifigure{\Fig\fga}{Schematic depiction of the classical boundary
trajectory for a sub-critical (left) and a super-critical (right)
shock wave.}{fga.eps}{2.50}

\subsec{Boundary trajectory for a shock-wave.}

The total system of matter and boundary only describes a well-defined
dynamical system if one restricts to field configurations below a
certain critical energy flux. As an example, we consider
the classical boundary equation
when the incoming wave is a shock wave located at $x^+ = q^+$,
with amplitude $p_+$
\eqn\shock{
T_{++}(x^+) = p_+ \delta(x^+-q^+).
}
As long as the total energy  $E=p_+q^+$ carried by the pulse is smaller
than ${m^2 \over 4\lambda^2}$, the boundary trajectory is time-like
everywhere and given by
\eqn\bounsh{ (\lambda^2 \X^-- p_+)\X^+ =
- {m^2 \over 4\lambda^2}
}
for $x^+ < q^+$ and
\eqn\bounsho{ \X^- (\lambda^2\X^++p_-) =
- {m^2 \over 4\lambda^2}}
with
\eqn\pmin{p_-= {\lambda^2 p_+ q^+
\over {m^2 \over 4\lambda^2q^+} - p_+}}
for $\X^+ > q^+$.
The typical form of this boundary trajectory is depicted in fig 1a.
Note that the mirror point indeed behaves as a particle with
negative rest mass: when the shock wave hits it, it does not bounce
back to the left but in the opposite direction to the right.

In case $p_+q^+ > {m^2 \over 4\lambda^2}$ then the solution to
the equation \bound\ cannot be time-like every where, but
turns space-like for $\X^+ > q^+$ (see fig 1b).
So in this regime it is no longer classically consistent to treat
the boundary as a reflecting mirror.
This example suggest that the following inequality
\eqn\ineq{P_+(x^+) < {m^2\over 4\lambda^2 x^+}}
for all $x^+$, with $P_+(x^+)$ defined in \ppdef, is a necessary and
possibly sufficient criterion for the energy flux to ensure that the
classical boundary remains timelike.
Note that this inequality does not imply any specific {\it local}
upper bound on the energy flux $T_{++}$.

\newsec {Classical Hamiltonian Formalism.}

To prepare for the transition to the quantum theory and to understand
better why the mirror-matter dynamics conserves energy, we will now consider
the reformulation of the system in hamiltonian language.
Our eventual goal will be to construct this hamiltonian in such a way that it
exactly describes the dynamics as seen by the asymptotic observer.

\subsec{Derivation of the hamiltonian.}

A hamiltonian formulation always depends on a specific choice of a space
and time coordinate. We will denote these by $\sigma $ and $\tau$,
respectively. To start with, it will be most convenient to
choose them in such a way that
the location of the boundary is always at a fixed value of $\sigma$.
This will have the advantage that we will not have to deal with a
dynamical restriction on the coordinates. Instead, the interaction
between the matter and the boundary $\X^\pm(\tau)$ will be described
via an explicit term in the action.
A particularly simple choice of coordinates $(\sigma,\tau)$ is as follows
\eqn\coordx{x^\pm(\sigma,\tau) = \X^\pm(\tau) \pm \sigma}
where the space coordinate is restricted to the positive
half line $\sigma >0$. The scalar field action then takes the following
form (here and in the rest of this section we will suppress the
(sum over the) $i$-index of the scalar fields $f_i$)
\eqn\actie{S_m = \int \! {d\tau d\sigma\over
\partial_\tau\X^+ + \partial_\tau\X^-}
 (\partial_\tau f - \partial_\tau \X^+ \partial_\sigma f)
(\partial_\tau f + \partial_\tau \X^-
\partial_\sigma f)}
The form of this action can be considerably
simplified, by introducing fields
$\pi_\sigma$ that denote the canonical conjugate of the $f$-fields. We can
then write the action in the first order form
\eqn\actieh{S_m = \int d\tau \int\limits_0^\infty \! d\sigma
(\pi_\sigma \partial_\tau f - \partial_\tau \X^+ T_{++} -
\partial_\tau \X^- T_{--})}
where
\eqn\stresst{T_{\pm\pm} = {1\over 4}(\pi_\sigma \pm \partial_\sigma f)^2}
denote the left and right-moving stress-tensors. The original action
\actie\ is recovered from \actieh\ by eliminating the fields $\pi_i$
via their
equations of motion.

The boundary action itself can also be written in a hamiltonian form
by introducing conjugate coordinates $\PP_\pm$ to $\X^\pm$ and
a lagrange multiplier field $e$
\eqn\actiebh{S_b = \int d\tau
(\PP_+ \partial_\tau \X^+ +  \PP_- \partial_\tau \X^-
+ \lambda^2 \X^- \partial_\tau \X^+ + e \ (\PP_+\PP_- + m^2))}
Again, this is the familiar form of the action of a
Klein-Gordon particle moving in a constant electric field
in two-dimensions.\footnote{*}{This point particle action was also
considered in relation with two-dimensional dilaton gravity in \cj.}

Both the scalar field action and the boundary
action are invariant under arbitrary
reparametrizations of the time variable $\tau$.
This local gauge invariance means that the hamiltonian, when constructed in
the standard way, is identically zero.
Indeed, by performing a redefinition of the momentum variables
$\PP_\pm$, the combined lagrangian can be written in the standard first
order form
\eqn\actiem{\eqalign{S = & \
\int \! d\tau \int\limits_0^\infty \!\! d\sigma\, \pi_\sigma \partial_\tau f
+ \int d\tau \, \PP_+\partial_\tau \X^+ + \PP_-\partial_\tau \X^-
\cr
 &
 \ \ \ \ \ +  \int d\tau \, e \, [(\PP_+ +{\lambda^2\over 2}\X^-- P_+)
(\PP_- - {\lambda^2 \over 2}\X^+ - P_-) + m^2 ]}}
with
\eqn\ppp{P_\pm = \int\limits_0^\infty \! d\sigma \, T_{\pm\pm}}
It is now manifest that we are dealing with a theory with a hamilton
constraint rather than a hamiltonian. This may look
somewhat surprising, since we started with a theory of free scalar fields
that contains states of non-zero energy. The reason is of course that
the dependence  of the physical coordinates $x^\pm$ on the fiducial
time $\tau$ is defined in \coordx\ via the arbitrarily
parametrized boundary trajectory $\X^\pm(\tau)$.

However, we want to construct a non-trivial
time-evolution and a corresponding
non-zero hamiltonian. To this end we will have to choose a specific
physical time $\T$ and fix the reparametrization invariance in $\tau$
by setting $\tau = \T$.
We would like this time coordinate to coincide
with the physical time as seen by an asymptotic observer. This leads us to
the following definition of $t$
\eqn\time{2\lambda \T(\tau) = \log(-\X^+/\X^-)}
We also define
\eqn\spc{2\lambda \R(\tau) = \log(-\lambda^2\X^+\X^-)}
Furthermore, instead of \coordx, we will choose a
parametrization of the original light-cone coordinates $x^\pm$ that is
appropriate for this choice of time
\eqn\coordy{\eqalign{x^\pm(\sigma,\tau)
& =  \X^\pm(\tau)e^{\lambda \sigma} \cr
& = \pm \lambda^{-1}
\exp[{\lambda(\pm \T(\tau) + \R(\tau) + \sigma )]}}}
with  $\sigma$ again defined on the positive half line.
For later reference, we note that the physical space coordinate $r$ is
equal to $\R(\tau) + \sigma$.
Now, after following the identical steps as above, we find
that in terms of these new variables the action takes the
form
\eqn\actien{\eqalign{
S = & \int \! d\tau \int\limits_0^\infty
\!\! d\sigma\, \pi_\sigma \partial_\tau f
+ \int d\tau \, \PP_\T \partial_\tau \T +  \PP_\R \partial_\tau \R
\cr
& \ \ \ +
\int d\tau \, e \, (-(\PP_\T - \lambda e^{2\lambda \R} - P_\T)^2
+  (\PP_\R - P_\R)^2
+ m^2e^{2\lambda \R})}}
with
\eqn\ph{P_\T = \int\limits_0^\infty \!\!
d\sigma  
{1\over 2}\,(\pi_\sigma^2 + (\partial_\sigma f)^2)}
\eqn\pp{P_\R
 =\int\limits_0^\infty \!\! d\sigma \, 
\pi_\sigma \partial_\sigma f}
This action is completely equivalent to the previous one,
and is again reparametrization invariant in the fiducial time $\tau$.
However, now we can identify a natural hamiltonian $H$ 
of the model,
namely as the operator
that generates the time translations in the {\it physical} time $\T$. From
the form \actien \ of the action we read off that
\eqn\ham{H 
= \PP_\T }
which, using the constraint imposed by $e$, can be solved
\eqn\ham{H 
= - \sqrt{(\PP_\R-P_\R)^2 + m^2 e^{2\lambda \R}} +
\lambda e^{2\lambda \R} + P_\T}
This hamiltonian acts in the phase space of the free fields
$(\pi, f)$ and of the boundary coordinates  $(\PP_\R, \R)$.

\subsec{Equations of motion and energy conservation.}

It is instructive to analyze the classical matter-mirror dynamics
in this coordinate system. From the Poisson brackets on the phase space
\eqn\poif{\{ \pi_\sigma(\sigma_1),f(\sigma_2)\} = \delta(\sigma_{12})}
\eqn\poir{ \{ \PP_\R,\R\} = 1}
and the form \ham\ of the hamiltonian, we deduce the following
equations of motion for the boundary variables $\R$ and $p_\R$
\eqn\reom{\dot{\R} = 
- {\PP_\R - P_\R \over
\sqrt{(\PP_\R-P_\R)^2 + m^2 e^{2\lambda \R}}}}
\eqn\peom{
{\dot{\PP}_\R} = {\lambda m^2 e^{2\lambda \R} \over
\sqrt{(\PP_\R-P_\R)^2 + m^2 e^{2\lambda \R}}}   - 2 \lambda^2 e^{2\lambda\R}}
with $\dot{\R} = {d\R \over d\T}$ etc.
The equation of motion for the $f$-fields can be written as
\eqn\eomf{\dot{f}= \pi_\sigma + \dot{\R} \, \partial_\sigma f}
\eqn\eompi{\dot{\pi}_\sigma = \partial_\sigma^2 f
+ \dot{\R} \, \partial_\sigma \pi_\sigma}
This can be integrated to the standard form of a sum of a left and
right-moving wave\footnote{*}{The suffices $in$ and $out$ here refer to
the respective asymptotic regions.}
\eqn\solf{f = f_{in}(\R(\T) \! + \sigma - \T) +
f_{out}(\R(\T) \! +\! \sigma\! +\! \T)}
and similar for $\pi_\sigma$. This shows that $\R(\T)$ indeed
parametrizes the physical location in $(r,t)$ space
of the mirror trajectory at $\sigma = 0$.

We can use the equation \reom\ to eliminate $\PP_\R$ from the
expression \ham\ for the hamiltonian. We find
\eqn\hamt{ H 
= - \gamma{m e^{\lambda \R}}
 + \lambda e^{2\lambda \R} + P_\T}
with
\eqn\gam{\gamma = {1\over \sqrt{1 - \dot{\R}^2}}}
The combination on on the right-hand-side of \hamt\ is constant in time and
can be identified with the total energy $E$ of the matter-mirror system.
Thus we find that the mirror behaves in this coordinate system as a
particle with a position dependent
negative mass $m(\R) = - m e^{\lambda\R}$ moving in electric field
$E(\R) = 2\lambda^2 e^{2\lambda \R}$.

To understand qualitatively
the motion of this particle, it is useful to write the force equation
in this language. The force due to the reflecting
matter is determined via the
rate of change in the total matter energy $P_\T$, given by
\eqn\ptdot{\dot{P}_\T = -(1+\dot{\R}) T_{in} + (1-\dot{\R}) T_{out}}
Here $T_{in}$ and $T_{out}$ denote the left-moving
$in$-flux and right-moving $out$-flux of stress-energy at the
location of the boundary. The two energy fluxes are related via
the classical reflection equation, which takes the form
\eqn\reflr{(1 +\dot{\R})^2T_{in} = (1-\dot{\R})^2T_{out}}
Thus we can express $\dot{P}_\T$ solely in terms of the incoming flux
$T_{in}$ as
\eqn\ptdott{\dot{P}_\T = \dot{\R} {\cal F}_m \qquad \qquad
{\cal F}_m = -2 {1-\dot{\R}\over 1+\dot{\R} } \, T_{in}}
The quantity ${\cal F}_m$ represents the force due to the matter bouncing
off the mirror.

The equation of motion of the mirror particle can
now be written in the following form
\eqn\reo{-\gamma^3 m e^{\lambda\R} \ddot{\R} = -{\partial V\over \partial \R}
+ {\cal F}_m}
with
\eqn\potent{V(\R) = - \gamma m e^{\lambda \R} + \lambda e^{2\lambda \R}}
The form of this potential energy is indicated in fig 2.

\ifigure{\Fig\fgb}{The form of the effective potential $V(\R)$,
showing the equilibrium position $\R = \R_0$. This equilibrium is
unstable because the effective mass of the mirror particle is negative.}
{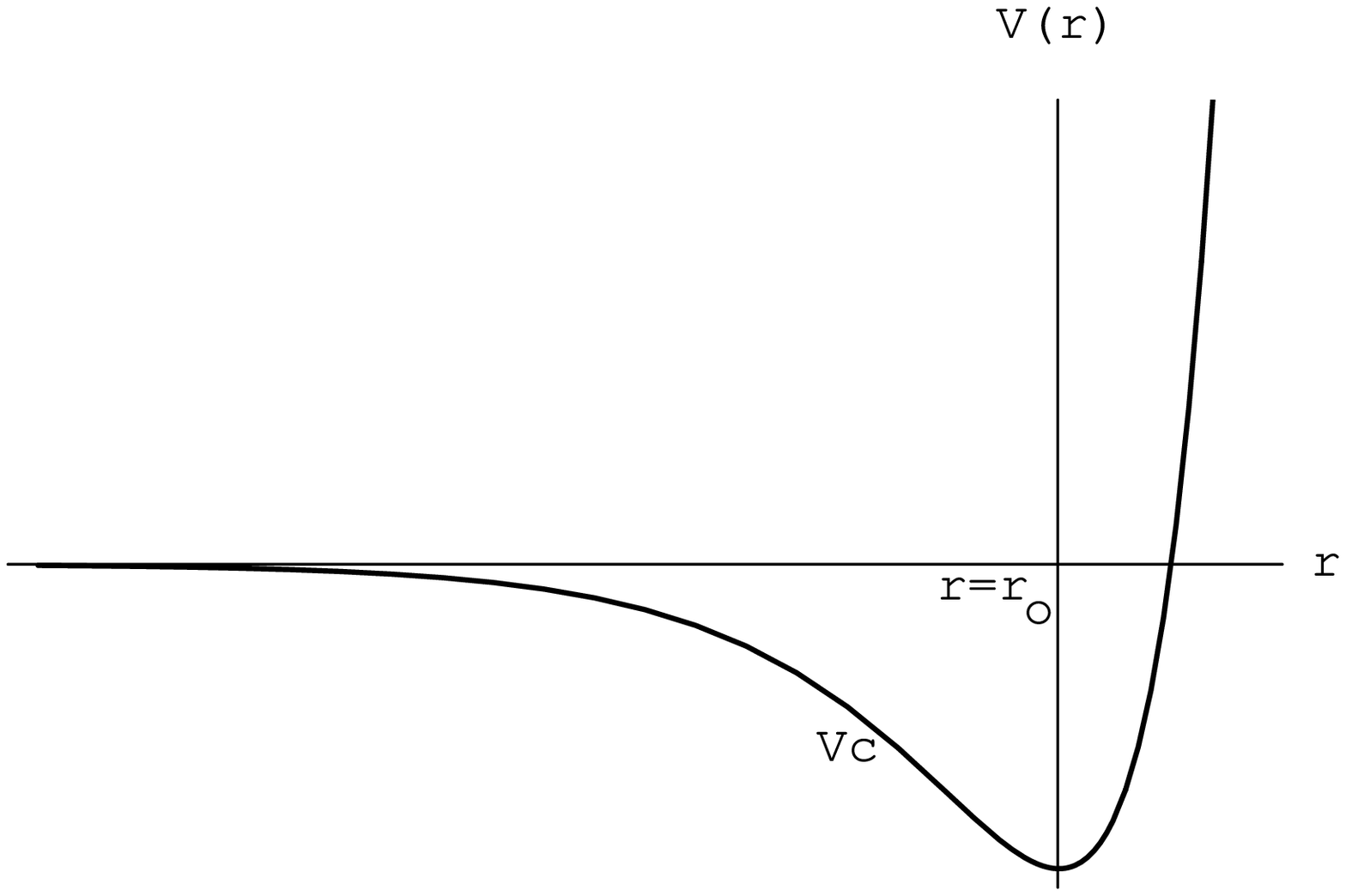}{1.75}

\subsec{Classical mirror trajectories.}

Before we describe what the motion of the mirror interacting with a matter
wave looks like,
let us first discuss the vacuum solutions to these equations, {\it i.e.}
with no matter present.
 From the shape of the effective potential we then see that there is
one isolated solution for which $\R$ is constant
\eqn\rco{e^{2\lambda \R_0} = {m^2 \over 4\lambda^2}}
This is the unique physical vacuum solution. Note, however, that
this trajectory is in fact unstable, due to the negative effective
mass of the boundary particle. The vacuum equations in principle
also allow for a more general class of solutions of the form
\eqn\rac{2 e^{\lambda (\R(\T)-\R_0)}
= \sqrt{4 + e^{2\lambda (\T-\T_0)}}
- e^{\lambda(\T - \T_0)} }
with $\T_0$ a constant of integration. They describe (see fig 3a)
a mirror that begins at $\R = \R_0$ and then starts to accelerate to the
left until it reaches a constant acceleration at late times and
approaches the light-like asymptote
\eqn\asymp{\R -\R_0 = -(\T - \T_0).}
The time reverse of this run-away solution is also a vacuum
solution.\footnote{*}{We should note that the
vacuum equations in fact also allow solutions
that run off to $\R = +\infty$ in a finite time. This motion,
however, is clearly unphysical and is eliminated after choosing appropriate
initial conditions.}

\ifigure{\Fig\fgc}{The form of the mirror trajectory in ($\R$,$\T$)
coordinates for a vacuum run-away solution (left) and the solution with
a subcritical matter shock wave (right).}{fgc.eps}{2.75}

The typical motion of the mirror interacting with a sub-critical
matter wave is now described as follows (cf. section 3.3).
A short time before the matter wave reaches the mirror, it
starts to pre-accelerate to the left
and follows one of the above vacuum trajectories of the form \rac.
The parameter $\T_0$ of this motion is not free but
determined in terms of the total $x^+$-momentum $P_+ = \int dx^+ T_{++}$
carried by the incoming $f$-wave via
\eqn\pc{P_+ = \lambda e^{\lambda(\R_0-\T_0)}}
If we consider the specific example of a matter pulse concentrated into
a shock wave along the line $r+t = 0$, with total energy $E$,
then this relation reads
\eqn\oc{E = \lambda e^{\lambda(\R_0 - \T_0)}.}
Provided the matter wave is subcritical, it will reflect off the mirror
in a finite time, after which the mirror will eventually return to the
(unstable) vacuum position $\R_0$ via a trajectory described by the
time reverse of \rac. The form of this trajectory is indicated in fig 3b.

It is clear what will happen in the super-critical regime.
For the shock wave example we deduce from \oc\ that, if the total energy
exceeds the critical energy
\eqn\ineq{E_c = \lambda e^{2\lambda \R_0},}
the asymptote will be located at $\R_0 + \T_0 <0$. Since the wave itself
travels along the line $r+t = 0$, the initial data in this case prevent
the incoming signal from ever reaching the mirror. This is the manifestation
in this coordinate system of black hole formation.
Classically, there will be information loss, in the sense that
the incoming wave will never be reflected into a right-moving signal.
The question we wish to address in the following sections
is whether this conclusion will continue to hold when we include
the quantum effect of particle creation and the corresponding
back reaction on the mirror trajectory.

Let us make one final important comment about the classical
model. From the above description of the
(sub-critical) motion of the boundary point
as seen by an asymptotic observer,
it may appear that we had to make an acausal fine-tuning \pc\ of the initial
data. This fine-tuning is necessary to ensure that after the reflection,
the mirror will eventually return to the original equilibrium point.
One should keep in mind, however, that it is not the mirror trajectory
but the $(r,t)$ coordinate system {\it itself} that is fine-tuned:
it is determined (causally) in terms of the mirror trajectory, via the
condition that the future asymptote corresponds to
$\R(\T) = \R_0$. In other words, the apparent acausality arises because
we describe the entire motion of the system in
terms of asymptotic coordinates
that depend on events taking place at late times. As we will discuss
in section 7, this fact will have important consequences at the
quantum level.

\newsec{Semi-Classical Analysis.}

We now wish to discuss the two-dimensional dilaton gravity model
at the semi-classical level. As pointed out in \cghs, one then
needs to include a one loop term in the action. This counter term
takes care of the conformal anomaly and also effectively represents the
back
reaction due to the quantum radiation emitted via the Hawking process.
The purpose of this section is to compute the effect of this counter term
on the equation of motion of the boundary. The final result for this
correction term will be the same as that described in section 2.

\subsec{Semi-classical dilaton gravity.}

It would greatly simplify the theory if we can choose the counter term
in a way that respects all the symmetries of the classical theory.
In particular, we would like to preserve the property
that the classical equations of motion imply that
the rescaled metric is flat everywhere, even in the presence of matter.
In addition, the counterterm is subject to the requirement that
asymptotically on ${\cal{I}}^+$ it must represent
the correct energy flux of the physical Hawking particles.
These considerations lead us to choose the following one-loop
correction to the effective action
\eqn\srs{
S_1=-{1\over 8\pi}\int d^2x[{N-24 \over 12}\hat{R}{1\over\nabla^2}\hat{R}-
    {N \over 12}(2\phi \hat{R})]}
with $\hat{R}$ being the Ricci scalar of the rescaled metric $\hat g$
defined in \resc. This choice of correction term
combines the approaches proposed in \rst\ and \str.
The first term above is the usual non-local quantum anomaly term that
compensates for the conformal anomaly, while
the second term is a local counterterm that ensures that in the far future
the physical metric $g_{ab}$ couples to the correct Hawking flux carried
by the physical particles.

Setting the conformal gauge,
\eqn\gauc{\eqalign{
g_{uu}=g_{vv}=&0 \cr
g_{uv}=g_{vu}=& - {1\over 2} e^{2\rho}
}}
the semi-classical effective action becomes
\eqn\a{\eqalign{
S={1\over\pi}\int d^2x [
  & e^{-2\phi}(2\partial_u\partial_v\rho -4\partial_u \phi
\partial_v\phi + \lambda ^2 e^ {2\rho} )
- {1\over 2} \sum_{i=1}^{N} (\partial_uf_i\partial_vf_i ) \cr
  & + 2 \kappa\phi\partial_u \partial_v (\rho-\phi)
- 2(\kappa-1)\partial_u (\rho-\phi)\partial_v (\rho-\phi)]
}}
with $ \kappa={N\over 24} $.
This one-loop effective
action can be brought into a form which is essentially identical
to the original classical action, by performing the change of variables
\eqn\b{
\eqalign{
\hat{\rho} =& \rho-\phi }}
\eqn\b{
\eqalign{
\Omega=& e^{-2\phi} + {\kappa}\phi
}}
The action then becomes
\eqn\svar{
S={1\over \pi} \int d^2x[
\lambda^2 e^{2\hat\rho} +
2(\Omega-\hat\kappa\hat\rho)
\partial_u \partial_v \hat\rho
+{1\over 2}\sum_{i=1}^N\partial_uf_i\partial_vf_i]
}
with $\hat\kappa = \kappa -1$. The
corresponding equations of motion of $\Omega$ and $\hat{\rho}$
\eqn\ha{
\partial_u \partial_v \hat\rho = 0}
\eqn\hb{
\partial_u \partial_v (\Omega -
\hat\kappa \hat\rho)
+ \lambda^2 e^{2\hat\rho} = 0}
indeed
take the same form as the classical equations (namely via the substitution
$ \Omega - \hat\kappa\hat \rho \rightarrow e^{-2\phi}$,
$\hat\rho = \rho - \phi$). The difference, however, between
	the one-loop corrected and the classical theory resides in the
form of the stress-energy tensor, which reads
\eqn\seta{
T_{uu} = -2  \partial_u \Omega \partial_u \hat\rho
+ \partial_u^2 \Omega  +  2\hat\kappa
(\partial_u \hat\rho \partial_u \hat\rho
- \partial_u^2 \hat\rho)}
and similar for $T_{vv}$.
Under the conformal transformations generated by these stress-tensors,
$\Omega$ transforms as a scalar, while $e^{2\hat\rho}$ defines a
(1,1) form.

The one-loop corrected dilaton gravity theory is still
essentially a free field theory. To see this, we notice
that the rescaled metric $\hat{ds}^2 = e^{2\hat\rho} du dv$
is still flat everywhere. As in the previous section, we can
in principle use this fact to perform a further gauge fixing, and
eliminate the conformal invariance by imposing the condition
$\hat{\rho} = 0$. Instead, however, let us for the moment keep
the conformal symmetry, and use the flatness of $\hat{ds}^2$
to introduce two chiral fields $X^\pm$ via
\eqn\hc{
\hat\rho(u,v) = {1 \over 2} \log(\partial_u  X^+(u)
\partial_v X^-(v))}
These fields parametrize the coordinate transformation from the $(u,v)$
system to the coordinates $x^\pm$ in which $\hat\rho =  0$.
The general solution to the equation of motion of $\Omega$ can
now be written as
\eqn\hd{
\Omega(u,v) = - \lambda^2 X^+(u) X^-(v) +
\omega^+(u) + \omega^- (v)}
The chiral fields $\hat{\rho}$ (or equivalently $X^\pm$) and $\omega_\pm$
are the free field variables. The classical Poisson brackets are
\eqn\poise{
\{\omega^+(u_1) , \partial_u \hat\rho (u_2)\}= {1\over 2} \delta (u_{12})
}
\eqn\poist{
\{\omega^+(u_1) , \partial_u \omega^+(u_2)\}=  {\hat\kappa}
\delta (u_{12})}
It is also useful to introduce variables $P_{\pm}$ via
\eqn\defa{
\partial_u \omega^+
= P_+ \partial_u X^+ + {\hat\kappa\over 2} \partial_u \log (\partial_u X^+)
}
\eqn\defb{
\partial_v \omega^-
= -P_- \partial_v X^- + {\hat\kappa\over 2} \partial_v \log( \partial_v X^-)
}
These variables $P^{\pm}$ are canonically conjugate to the coordinate
fields $X^{\pm}$,
\eqn\canon{
\{P_{\pm}(u_1) , \partial X^{\pm}(u_2)\} = \delta (u_{12})}
while the $P_\pm$ fields commute among themselves. This last requirement
fixes the form of the second term in \defa\ and \defb.
The stress-tensors in the new variables read
\eqn\tena{
T_{uu} = \partial_u X^+ \partial_u P_+
- {\hat\kappa\over 2} \partial_u^2 \log \partial_u X^+
}
\eqn\tenb{
T_{vv} = -\partial_v X^- \partial_v P_-
- {\hat\kappa\over 2} \partial_v^2 \log \partial_v X^-
}
These satisfy the Poisson bracket of a Virasoro algebra with
central charge $\hat\kappa$. Note that the $P_\pm$ fields do not have
the simple scalar transformation law under the conformal symmetry
generated by this stress-tensor, but transform such that the combination
on the right-hand side of \defa\ and \defb\ behave as proper conformal fields
of dimension 1.

In the remainder of this section
we will restrict our attention to the semi-classical
physics of this system. In principle, this can be justified only if we
take the limit of large $N$. For this reason we will no longer make
any distinction between $\hat{\kappa}$ and $\kappa$, as they become
identical in this limit.

\subsec {Boundary conditions.}

Let us return to the discussion of the boundary conditions.
We choose the $(u,v)$-coordinate system in such a way that
the boundary becomes identified with the line $u=v$, and denote
the parameter along this boundary by $s$.
As suggested above, we
first require that the dilaton field is
constant along the boundary
\eqn\com{\partial_s \Omega = 0}
where $s = u =v$ is the coordinate along the boundary.
In terms of the $X$ and $P$-fields this
condition reads
\eqn\bon{
\partial_s X^+ (\lambda^2 X^- - P^+) + \partial_s X^- (\lambda^2 X^+ + P^-)
+ {\kappa\over 2} \partial_s \log(\partial_s X^+ \partial_s X^-) = 0.}
The above boundary
condition is coordinate invariant, which
allows us to impose the additional condition that the
gravitational and matter components of the energy momentum
flux each separately get directly reflected off the boundary. So
the equation \bon\ is supplemented with the condition that
\eqn\hct{T^{g}_{uu} =T^{g}_{vv} }
with $T^g_{uu}$ and $T^g_{vv}$ as given in \tena\ and \tenb.
The two equations \bon\ and \hct\
combined specify the precise reflection condition
that relates the incoming canonical variables $(X^+,P_+)$ to the
outgoing canonical variables $(X^-,P_-)$.

We would like to make manifest
that this relation defines a canonical transformation.
To this end, we should write a generating function $S[X^+,X^-]$
of the coordinate fields such that the momenta $P_\pm$ defined by
\eqn\gena{
P_{\pm} = {\delta S[X] \over \delta \partial_s X^{\pm}}
}
identically solve the boundary equations \bon\ and \hct.
This results in a set of functional equations for $S[X]$
that can be solved explicitly. The form of the solution is
unique, once we fix the constant value of $\Omega$ along the boundary.
If we set $\Omega(\X^+,\X^-) = {m^2 \over 4\lambda^2}$,
then the generating functional $S[X]$ takes the following form
\eqn\genfun{\eqalign{
S[X] = & m \int ds \sqrt{\partial_s {X}^+ \partial_s{X}^-}
- \lambda^2\int\! ds \,X^+\partial_s {X}^- \cr
& \ \ \ \ \ \ \ + {\kappa\over 2}
\int ds \log(\partial_s X^+) \partial_s \log(\partial_s X^-)}}
In this result we recognize
the classical boundary action discussed in section 3, and the above formula
can be interpreted as the quantum corrected version of it.
The formulas \gena\ defining the momenta $P_\pm$ become
\eqn\newmom{
\pm {m\over 2}  \sqrt{\partial_s X^+ \partial_s X^-} +
\partial_sX^\pm(\lambda^2 X^\mp \mp P_\pm) + {\kappa \over 2} \partial_s
\log(\partial_s X^\mp) = 0}
In the next subsection we will see that this relation between the $in$-
and $out$-dilaton gravity fields
can be reinterpreted as the semi-classical equations
of motion that determine the boundary trajectory for given $in$ and
$out$ energy flux.

The action \genfun\ and the relations \newmom\ are
well-defined as long as both $X^+(s)$ and $X^-(s)$ are invertible
functions of $s$. This, however, puts a non-trivial
restriction on the possible values of the canonical momenta $P_\pm$,
which, as we will see shortly, is related to the inequality
\ineq\ on the incoming matter energy flux. For the time being,
we will restrict our attention to this sub-critical regime.

\subsec{The physical mirror trajectory.}

We are now finally in a position to derive the semi-classical
correction to the classical matter-mirror dynamics
described in section 3 and make
contact with the discussion in section 2.
To do this we must eliminate the redundancy due
to the conformal invariance and
translate the above description of the scattering
off the boundary into physical variables that commute with physical constraint
\eqn\constraint{T^g_{uu} + T^m_{uu} = 0.}
 The simplest way to do this is
to use the physical fields $X^\pm$ as the reference coordinate system.
This procedure amounts to choosing the gauge $\rho=\phi$.
Thus we are instructed to adopt
the following definition of the physical left-moving field
$f_i(\X^+)$
\eqn\fphys{f_i(\X^+)= \int \!\! du  \partial_u X^+
\delta(X^+(u) - \X^+) \, f_i(u) }
where $\X^+$ denotes a $c$-number coordinate.
The same construction can be used to define the physical fields
$f_i(\X^-)$ in the left moving sector.
Furthermore, we can also introduce physical variables
$\hat{P}_\pm(\X^\pm)$ that are made from the gravitational fields via
\eqn\pphys{
\hat{P}_+(\X^+) = \int \!\! du \, (P_+ \partial_u X^+
+ {\kappa \over 2} \partial_u\log (\partial_u X^+))
\, \delta(X^+(u) - \X^+)}
Here the combination on the right-hand side has been chosen such that
it commutes with the gravitational stress-tensor $T^g_{uu}$.
\footnote{*}{Note
that, although we assume that $X^+(u)$ is invertible, the above
definitions of the physical variables would in principle also make sense
without this restriction.}

The physical fields $\hat{P}_\pm(\X^\pm)$ are not independent from the
matter fields, but are related to them via the Virasoro constraint
equations \constraint and identified with the integrals
with
\eqn\ppdef{\hat{P}_\pm(\X^\pm)
= \pm \int\limits_{{\X^\pm}}^{\pm\infty} \!\!
dx^\pm\, T_{\pm\pm}.}
of the respective components of the matter stress-tensor.
Thus we can now finally determine the physical trajectory $\X^-(\X^+)$ of
the boundary by rewriting the relations \newmom\ in terms of the
physical variables $\hat P_\pm(\X^\pm)$. These relations give a
semi-classical equation of motion, which takes the form
\eqn\eoma{
-{m\over 2}\sqrt{\partial_\pm \X^\pm} \pm \lambda^2 \X^\pm +
\hat P_\mp(\X^\mp)
\pm{\kappa\over 2}\partial_\mp \log \partial_\pm \X^\pm = 0
}
These are the quantum corrected conservation laws of total
$\X^\pm$-momentum. These differential equations are
(upto the sign of $m$) identical to the ones we arrived at in section 2.
There we wrote them in terms of the proper time
coordinate $\tau$ along the boundary, as defined in \fix, as
\eqn\momk{
-{1\over 2} m {\partial_\tau \X}^\pm  \pm {\kappa} {\partial_\tau^2
\X^\pm}
\pm \lambda^2 \X^\pm +  \hat P_\mp(\X^\mp) = 0
}
As was shown before, these semi-classical momentum conservation
laws are compatible with the
the quantum reflection equation \reflq\ for the stress-energy tensor.
In the following subsection we will
investigate some of the physical properties of these equations.

First let us determine the range of parameters for which there still
exist vacuum trajectories of the form
\eqn\vact{-\lambda^2 \X^+\X^- =
e^{2\lambda \R_0}.}
For this we must take into account that
the stress-energy flux $T_{\pm\pm}$ is in fact not identically zero in the
vacuum. The physical vacuum is defined with respect
to the physical asymptotic time $t$, while $T_{\pm\pm}$ is normal ordered
with respect to the $x^\pm$ coordinate system. Due to conformal anomaly,
this implies that  $T_{\pm\pm}$ receives a negative contribution equal
to $-\kappa/2(x^+)^2$.  Thus we find that the vacuum trajectory \vact\
solves \eoma\ when $\R_0$ is a solution to
\eqn\eqq{2\lambda e^{2\lambda \R_0} - me^{\lambda \R_0}
+ {\kappa} \lambda = 0.}
In order for the solution to be real, the parameter $m$ has to be chosen
such that
\eqn\ineqq{m^2 > 8 \kappa \lambda^2.}
Hence, to have a time-like vacuum boundary trajectory, we
can no longer take $m$ to be arbitrarily small. Note further that,
if \ineqq\ is satisfied, equation \eqq\ has in fact {\it two} real
solutions.

\newsec {Semi-Classical Hamiltonian Dynamics.}

We now wish to return to the issue of energy conservation in this model,
while including the quantum radiation and the effect of the back reaction.
Let us therefore again translate the above equations of motion
into the $(\R,\T)$ coordinate system. In section 4 we did this
via the hamiltonian formulation. This method, however, is now
not so easily available, because of the higher order derivatives
present in the quantum corrections. Still, the time-translation
symmetry guarantees that there exists an expression for the
conserved energy, that generalizes the classical result \hamt.

\subsec {Semi-classical energy conservation.}

The quantum corrected energy equation can be derived directly from the two
boundary equations \momk\ as follows. The total
energy $P_t$ and the total $\X^\pm$-momenta $P_\pm$ are related by
\eqn\entrans{
P_\T(\tau) 
= \lambda \int^\tau d\tau
\Bigl[ \X^+ \partial_\tau P_+ - \X^-  \partial_\tau P_- +
{\kappa\over 2} \partial_\tau \log(-\lambda^2 \X^+\X^-)\Bigr]}
The term proportional to $\kappa$ arises due to the anomalous transformation
law of the stress energy tensor.
Now we use the momentum conservation equations \momk\
 to eliminate the matter momenta on the right-hand side of \entrans\
in favor of
the ($\tau$-derivatives) of the boundary coordinates $\X^\pm$, which we
in turn replace by $\R$ and $\T$.
After some algebra, this procedure leads to the following equation
\eqn\tdot{-m e^{2\lambda \R} \partial_\tau\T +
2 \kappa e^{2\lambda \R}[\partial^2_\tau \R
+ 2\lambda(\partial_\tau \R)^2] +
\lambda e^{2\lambda\R} + {\kappa}\lambda^2 \R + P_\T   = {\rm constant}}
The constant on the right-hand side is the total conserved
energy.

We can eliminate the time-variable $\tau$ in favor of the physical
time $\T$, by writing equation \fix
\eqn\dtdtau{\qquad \qquad \partial_\tau \T =  \gamma e^{-\lambda \R}
\qquad \qquad
\gamma = {1\over \sqrt{1 - \dot{\R}^2}}}
with $\dot\R = {d\R \over d\T}$.
This allows us to remove $\tau$ from the
energy equation \tdot, leading to the following semi-classical
result for the total energy
\eqn\hamq{ H 
= -\gamma m e^{\lambda \R} + 2\kappa\lambda \gamma^2+
2 \kappa\gamma^4 \ddot{\R}
 + \lambda e^{2\lambda \R} + \kappa \lambda^2 \R +P_\T}
We want to use this expression to examine the energy
balance and (in)stability
of the mirror-matter dynamics.

As before, we can try to obtain a qualitative understanding of the boundary
dynamics by considering the force equation of motion.
This equation is obtained
by computing the time derivative of \hamq\
and putting the result equal to zero after dividing by an overall factor of
$\dot{\R}$. The time-derivative $\dot{P}_\T$ of the total matter
energy is still given by the classical expression \ptdot\ in terms of the
left- and right-moving energy flux. The reflection equation \reflr, however,
now receives an extra quantum contribution
\eqn\ptf{(1+\dot{\R})^2 T_{in}=(1-\dot{\R})^2T_{out}
+ 2\kappa \gamma^2 (\R^{(3)} + 3 \gamma^2 \dot{\R} \ddot{\R}^2)}
with $r^{(3)} = {d^3 {\R} \over {dt}^3} $.
Thus the change in the total matter energy in terms of the
incoming flux now becomes
\eqn\ptf{ \dot{P}_\T  =  2 \dot{\R}
 {1+\dot{\R}\over 1-\dot{\R}} T_{in}
- 2\kappa  \gamma^4  (1+\dot{\R})
(\R^{(3)} + 3 \gamma^2 \dot{\R} \ddot{\R}^2)}
The second term represents the change in the total matter energy due to the
quantum radiation by the accelerating mirror.

The final result for the quantum equation of motion of the mirror reads
\eqn\for{
\gamma^3m_q(\R) \ddot {\R}= -{\partial V_q \over \partial \R}
+ {\cal F}_m +  {\cal F}_q}
where
\eqn\qpot{\eqalign{m_q(\R) & = - m e^{\lambda \R} + 4 \kappa \lambda^2
\gamma \cr
V_q(\R) & = - \gamma
m e^{\lambda \R} + \lambda e^{2\lambda r} +
\kappa\lambda \R \cr
{\cal F}_m & = - 2{{1+\dot{\R}} \over 1- \dot{\R}} T^{in}\cr
{\cal F}_q & =  2\kappa \gamma^4  r^{(3)}
- 2\kappa\gamma^6(1-3\dot{\R}) \ddot{\R} ^2
}}
Here $m_q(\R)$ and $V_q(\R)$ are the quantum corrected effective
mass and potential energy of the mirror particle.
The shape of the potential $V_q(\R)$ is plotted in fig 4.
The term ${\cal F}_m$ is recognized as the classical force due to the
direct
reflection of the matter waves and ${\cal F}_q$ is the remaining part of
the force induced by the quantum anomaly. The sum of all the terms
proportional to $\kappa$ represent the total back reaction due to
the quantum radiation.

\ifigure{\Fig\fgd}{The form of the quantum effective potential $V_q(\R)$,
showing the equilibrium position $\R = \R_0$ and a second
extremum. For the run-away solutions described in section 6.2,
the difference between the
quantum and classical potential energy is exactly cancelled by
energy contained in the emitted quantum radiation.}{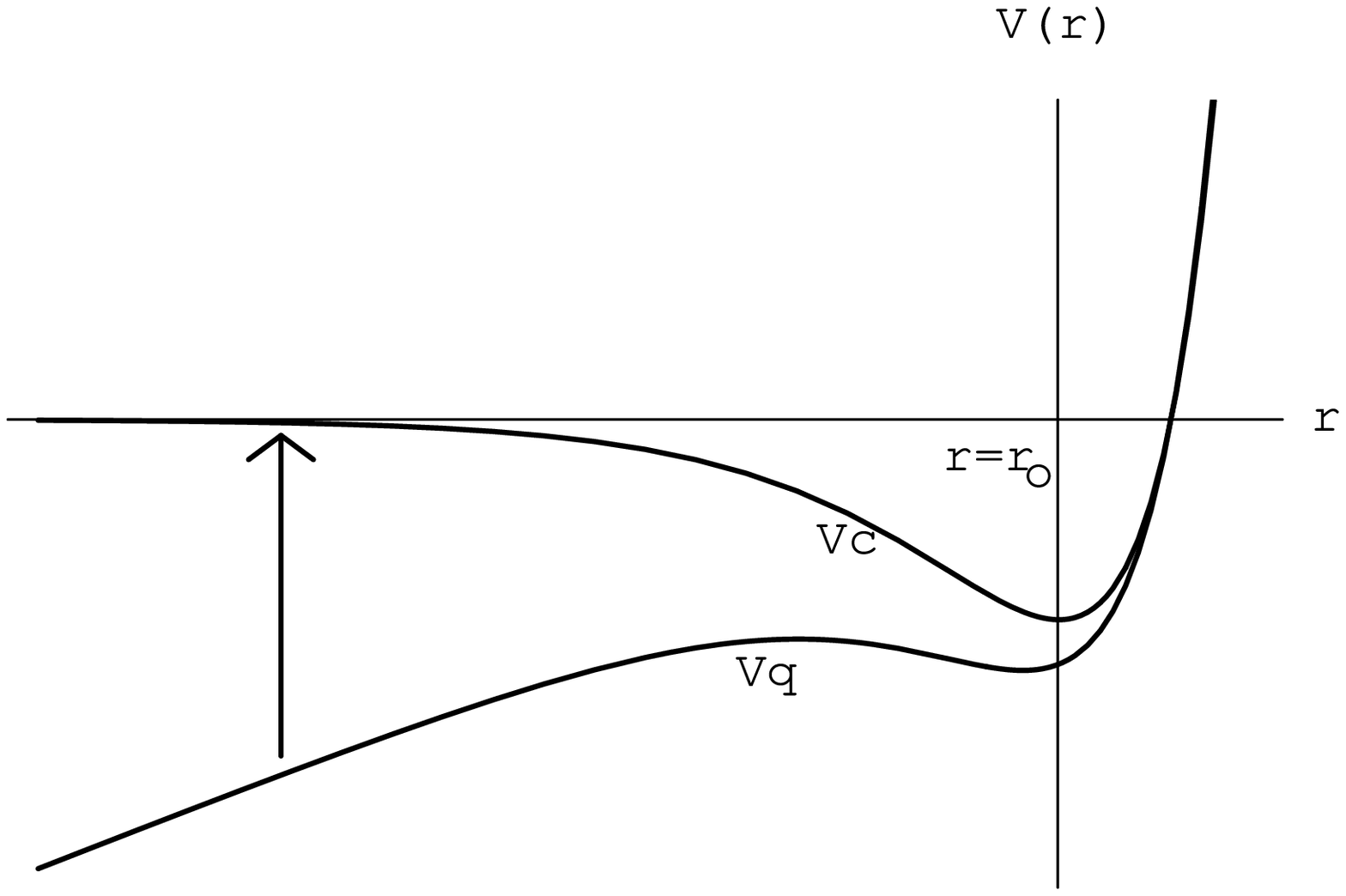}{2.25}

\subsec{Instability of the semi-classical equations.}

What do the semi-classical boundary trajectories look like? First, we
note that there still exists a vacuum trajectory of the form
$\R = \R_0$. Here $\R_0$ is found by solving
$\partial V_q/\partial \R = 0$, which is the equation \eqq.
The interesting question, however, is whether the semi-classical vacuum
equations of motion still allow for run-away solutions of the form
\eqn\racs{2 e^{\lambda (\R(\T)-\R_0)}
= \sqrt{4 + e^{2\lambda (\T-\T_0)}}
- e^{\lambda(\T - \T_0)} ,}
that describe a mirror that keeps accelerating to the left (see fig 3a).
This is not immediately obvious,
since the mirror will now radiate and produce a constant
flux of energy. From \ptf\ (with $T_{in}$ put to zero)
we can explicitly evaluate the energy as a function of the mirror
position $\R(\T)$ for the trajectory \racs
\eqn\enflux{P_\T(\T) = \kappa \lambda^2 (\R_0-\R(\T)) -
{\kappa \lambda \over 2} e^{-2\lambda(\R_0-\R(\T))}.}
This expression
indeed grows linearly for large $\T$.

Where does this energy come from? The only possibility appears to be
that the energy carried by the mirror itself grows negative at the same rate.
The semi-classical hamiltonian \hamt\ indeed contains several terms that can
become negative, in particular the term $\kappa \lambda^2 \R$ in the
effective potential energy $V_q(\R)$. As indicated in fig 4, the linear
growth \enflux\ in the energy $P_\T$ cancels against this linear term
in $V_q(\R)$ and effectively restores the classical form of the potential.
As a result of this cancelation, we find that
the total energy for the run-away solution \racs\ is indeed constant,
provided $\R_0$ solves \eqq. Thus we conclude that \racs\ is still
an allowed solution of the semi-classical vacuum equations of motion.

Let us now consider the semi-classical modifications to classical
mirror-matter dynamics described in section 3.3. At early times, there
is not too much difference. Some time before the matter wave arrives,
the mirror will start to pre-accelerate to left along the trajectory
\racs, where the parameter $\T_0$ is determined in the same way as before
in terms of the total $x^+$-momentum of the incoming matter wave via
\eqn\pcs{P_+ =  \lambda e^{\lambda(\R_0-\T_0)}}
except now with the new value for $\R_0$. Here
\eqn\pk{P_+ = \int\limits_{-\infty}^\infty
\! dt \, e^{-\lambda(r+t)} T_{in} (r+t).}
So there will still exist a
sub-critical regime in which the boundary trajectories will look roughly
as in fig 3b, and for which the above equations in principle provide an
explicit and well-defined relation between the in- and out-going waves.
This low energy scattering equation is energy and information preserving,
even though it is possible that a sizable fraction of the out-going
matter comes out in the form of thermal radiation emitted during the
period of pre-acceleration.

In the super-critical regime, on the other hand, we seem to find that,
in spite of the presence of the quantum radiation, the semi-classical
model has the same instability as the classical theory. It still seems
true that when the matter shock wave carries more than the critical energy
$E_c = \lambda e^{2 \R_0}$, the mirror point will keep accelerating
Forever and the matter pulse will never catch up with it.
So there will still be information loss, since eventually the incoming
wave will end up in the left asymptotic region and disappear into the
black hole.

However, unlike the classical case, there is now something
very unphysical about this situation, because if the mirror never stops
accelerating, it also never stops emitting radiation.
Clearly, in any physically realistic model, a black hole should
stop radiating and disappear after it has emitted all of its mass.
In our model, on the other hand, it seems that the energy of the mirror can
become arbitrarily negative, so no mechanism seems available that
stops it from accelerating after all in-coming energy has been emitted.
The question is how to deal with this (apparent) instability?

\newsec{The Super-Critical Case.}

The situation we have arrived at is similar to the instability that
at some point seemed to plague other semi-classical models of
two-dimensional dilaton gravity (see e.g. \bical). A particular
prescription for stabilizing the system has been proposed in \rst.
When translated into the language of our moving mirror model,
this proposal effectively amounts to imposing the boundary
condition that, when all the energy has been emitted, the mirror
jumps back to the original equilibrium point $\R = \R_0$, after which
it stops radiating. This prescription mimics the physical effect of
the black hole completely disappearing into nothing after it has
evaporated. By imposing this boundary condition, it seems one has indeed
restored the stability of the system, but only at a very high cost.
At the instant the mirror jumps back, all part of space to
the left of $\R=\R_0$ suddenly becomes forever invisible to the
outside world and the corresponding information can not be
retrieved. Thus, based on this physical picture, information loss
seems inevitable, at least at the level of this
semi-classical approximation.

We believe, however, that this conclusion is drawn too quickly.
The reason is that, as we will now argue, the
instability described in section 6.2
may well be an artifact of the approximations
that have been made. Namely, an important point that has not
yet been properly taken into account is that the parameter $\T_0$ in the
run-away solution \racs\ is not a $c$-number, but an operator that
depends on the incoming $x^+$-momentum via \pcs.  This equation is
directly related to the fact that the $(r,t)$ coordinates are determined
dynamically in terms of the incoming matter flux. Because this
relation is non-local, it leads to some
surprising consequences at the quantum level. In particular, we will
argue that it opens up the possibility for an alternative
physical mechanism for shutting down the radiation by the mirror,
that does not destroy quantum coherence.

\subsec{The algebra of $in$ and $out$-fields.}

The most direct and important consequence of the non-local
initial condition \pcs\ of the mirror trajectory
in $(r,t)$ space is that the left- and right-moving field operators
$f_{in}(r_1,t_1)$ and $f_{out}(r_2,t_2)$ will
{\it not} commute with each other.
Instead they will satisfy a non-trivial commutator algebra,
even when they are space-like separated.
The physical interpretation of this non-local algebra is that it represents
a gravitational shock-wave interaction between the incoming and out-going
matter waves. As has been emphasized by `t Hooft in the context of
$3+1$-dimensional black holes \thooft, whenever one sends in a particle
into a black hole, it will produce a gravitational shock wave near the
horizon, that results in an exponentially growing shift in the
position of all out-going particles.
In the present model, this gravitational shock-wave is generated via
the small shift in the mirror trajectory as a result of the small change in
the parameter $\T_0$ due to the $x^+$-momentum of the in-falling particle.

To compute the resulting algebra,
let us consider the semi-classical run-away solution \racs, and use it
to express the
$out$-fields $f_{out}$ in terms of the $in$-fields $f_{in}$.
Thus we imagine sending some
test-wave back in time and letting it reflect off the
space-time trajectory of the mirror point. This leads to
the relation
\eqn\scat{f_{out}(t-r) =
f_{in}(\R_0 + \T_0 - {1\over \lambda} \log[1+ e^{\lambda
(r-\R_0 - t+ \T_0)}])}
Although in the full quantum theory
this equation 
is not entirely exact (since it does not include all effects
of the back reaction
due to the test-wave itself) it should be accurate for $in$ and $out$
waves of reasonable energies and it becomes exact in the semi-classical
limit when $\hbar \rightarrow 0$.

Once we adopt this approximation, then
we immediately note that the above relation looks non-invertible, since
the argument of $f_{in}$ never exceeds the limiting value $\R_0+\T_0$.
This corresponds to the classical fact that $f_{in}$ waves that depart
at any later time will never be reflected to $out$-waves.  Now,
if we would treat $\T_0$ just as
a $c$-number quantity, we would conclude from this that the commutator
\eqn\false{
[f_{out}(r_2,t_2), f_{in}(r_1,t_1)]}
will vanish in the region
\eqn\limit{r_1 + t_1 > \R_0 + \T_0.}
If this were indeed true, then this would prove that the
$in$ Hilbert space is larger than the $out$ Hilbert space, and
this would imply that information loss is inevitable.
Instead, however, since $\T_0$ is expressed in terms of the total
in-coming $x^+$-momentum via the relation \pcs,
we must take into account that there is a non-zero commutator between
$\T_0$ and $f_{in}(r_1,t_1)$. A simple computation gives
\eqn\tcomf{[ \T_0, f_{in}(r_1,t_1)] =
i \lambda^{-2}
e^{-\lambda(r_1+t_1+\R_0  -\T_0)} \partial_{r_1} f_{in}(r_1,t_1)}
Although the right-hand side
becomes exponentially small at late times $t_1$,
its effect in \false\ can become very large.
After combining \tcomf\ with \scat, we can now compute the commutator
\false, with the result
\eqn\fcomf{[f_{out}(2), f_{in}(1)] =
i \lambda^{-2} e^{\lambda(-r_1-r_2 + t_2- t_1)}
\partial_{2}f_{out}(2) \,
\partial_{1}f_{in}(1)}
The above expression
for the algebra is valid in the semi-classical limit
$\hbar \rightarrow 0$ and then only in the regime \limit, which is for
those incoming waves $f_{in}$
that would classically never reflect off the mirror trajectory \racs.
For $r_1+t_1 < \R_0+\T_0$ there will be other contributions since
the left- and right-movers can then interact directly via reflection
off the mirror. Note further that this algebra is {\it symmetric} between
$f_{in}$ and $f_{out}$, although its derivation looked asymmetric.

\ifigure{\Fig\fge}{Due to the non-local definition of the $(r,t)$ coordinate
system, the left- and right-moving fields do
{\it not} commute with each other, even when they are space-like
separated. }{fge.eps}{2.25}

For correctness  we
should note that, since
the right-hand side  of \fcomf\ grows exponentially with the
time-difference $t_2-t_1$ and also for negative $r_1+r_2$,
it is after some point no longer a good approximation to work to first
order in $\hbar$. For large $t_2-t_1$
this semi-classical approximation breaks down
and we must start to take into
account multiple commutators between $\T_0$ and $f_{in}$. The net
result of this is to replace the classical algebra \fcomf\ by a
quantum exchange algebra of the form
\eqn\fexf{f_{out}(2) f_{in}(1) =
\exp\Bigl[\lambda^{-2} e^{\lambda(-r_1-r_2 +t_2- t_1)}
\partial_{1}
\partial_{2}\Bigr] f_{in}(1)f_{out}(2)}
This exchange algebra reduces to the semi-classical
formula in the $\hbar \rightarrow 0$ limit. It
explicitly reveals the shock-wave interaction
between the left and right-movers: it shows that when an
$in$ and $out$-wave cross, each will undergo an exponentially growing
displacement proportional to the $x^\pm$-momentum carried by the
other wave (see \svv).

In the following, however,
we will continue to work in the semi-classical limit.
It is clear, however, that even
in this case the non-local algebra \fcomf\ has non-trivial
consequences. In particular, it tells us that, due
to the quantum uncertainty principle, we should be very careful in
making simultaneous statements about the left- and right-moving fields
(that is, as long as we work in the $(r,t)$ coordinate system).

\subsec {The super-critical energy balance.}

A second important consequence of the relation \pcs\ is that
it turns \scat\ into a non-linear,  {\it energy preserving} relation
between the $in$- and $out$-fields. Namely, we can use the
conserved quantum hamiltonian $H$, that generates the time-evolution
of the $in$- and $out$-fields via
\eqn\hfin{ -i \partial_t f_{in,out}(t\pm r) = [H,f_{in,out}(t\pm r)] }
to compare the energy carried by the $out$-wave
(the left-hand side of \scat)
to that carried by the same wave before reflection off the mirror
(the right-hand side of \scat). If we would treat $\T_0$ as
a $c$-number, these energies would clearly be different. Indeed,
since the mirror recedes fast to the left,
the reflected $out$-signal has a much lower frequency than the
$in$-signal. In our dynamical mirror model, on the other hand,
any energy change must be compensated by an energy change
somewhere else. To make this
energy balance work, we must again take into account \pcs.
 From this relation we find that, as an operator in the $in$-Hilbert
space, $\T_0$ satisfies
\eqn\tham{[H , \T_0] = i.}
Here we used that $[H, P_+] = i\lambda P_+$, which holds formally.
It is now a simple calculation to show that the commutators of
$H$ with the left- respectively right-hand side of \scat\ are
indeed equal. Thus, as quantum operators, the expressions
on both sides carry the same energy.

Physically this means that, when we send
back a signal from the $out$-region and let it reflect to
an $in$-signal, the increase in energy due to the blue-shift
will be exactly compensated by a decrease in the energy carried by
the matter forming the black hole. The interaction
responsible for this non-local transfer of energy is the just
described commutator algebra \fcomf.
Indeed, another way of seeing that the super-critical scattering
equation \scat\ preserves energy is that the resulting algebra \fcomf\ is
time-translation invariant, and thus also energy preserving.

This observation again teaches us a useful lesson.
Namely, it tells us that it is essentially {\it impossible}
that more energy is contained in the out-going radiation than went in,
and thus that the instability that seems to plague the
semi-classical model should disappear in a proper quantum treatment of
the matter-mirror dynamics. We will now use this important insight,
in combination with non-local algebra \fcomf, to propose a new
physical mechanism that will indeed stop the acceleration of
the mirror by the time the total $in$ and $out$ energy are equal.
This should then automatically restore the stability of the model.

\subsec{Effective time-evolution of the in-falling matter.}

Let us now reconsider the time-evolution of the matter-mirror system
in the light of these new insights. To this end, let us
imagine setting up an experiment along a time slice $t= constant$
in which we try to detect all the in- and out-going particles and
measure their positions and energies.\footnote{*}{Warning: This
thought experiment is in fact highly hypothetical, since $r$ and $t$
have only real physical meaning as an asymptotic coordinate system.}
Due the non-local commutator between both kinds of particles,
we can do this experiment only upto a certain accuracy.
Mathematically, the Hilbert space of the $f$-fields
at a given time $t$ does not decompose into a simple tensor
product of a left- and right-moving Hilbert space, since it
makes a difference if we put, say, all out-going operators
$f_{out}$ to the left or to the right of the in-falling fields $f_{in}$.
It is therefore an ill-defined question how much
energy at a given time is contained in the in-falling or the out-going
matter. The answer depends on the specific ordering prescription.
Only the total (left plus right) energy remains well-defined.

It is now clear that the description we gave in section 6 needs
to be modified. There we had assumed that we could
assign simultaneous physical meaning to the energy carried by the
left- and right-moving fields. This allowed us to split the total
hamiltonian $H$ into two parts, a part $H_R$ that measures the
energy contained the out-going quantum radiation, given by \enflux,
and a part $H_L$ that measures all the remaining energy, given by
\eqn\hleft{H_L(\T) = H - \kappa \lambda^2 (\R_0-\R(\T)) +
{\kappa \lambda \over 2} e^{-2\lambda(\R_0-\R(\T))}}
In section 6 we interpreted
this expression as the constant total energy $H$ of the left-moving
matter plus the negative (quantum) effective potential of the mirror.
Our new proposal, however, is that this negative potential energy
does in fact not really exist, but has its origin in
the uncertainty relation between the left- and right-moving
energies. It is namely far from clear that the ordering
prescription that gives rise to the result \enflux\ for the
out-going energy is the same as the ordering in which the
in-going energy remains constant. In fact, intuitively one expects
these orderings to be opposite to each other, since to measure
the out-going energy one needs to bring all ${out}$-fields to the left of
the $in$-fields and the other way around if one wants to measure the
total in-going energy. Thus, if this intuition is correct, it means
that, in assigning simultaneous physical reality to the energy $H_R$ of the
Hawking radiation and the constant energy $H$ of the in-falling matter,
we have in fact
over-estimated the total amount of energy carried by the matter.
The super-critical energy balance described above suggests that
correcting the mistake should in fact precisely cancel the negative quantum
contribution to the potential energy of the mirror.

Thus we arrive at a new physical picture in which there is
a certain {\it complementarity} between the physical realities
as seen by an asymptotic observer and that seen by an in-falling
observer \thooft \comp. The reason for this complementarity
is that each of these observers will use a
different ordering prescription to assign physical meaning to the
same quantum state. For the in-falling observer, the in-falling
matter will simply propagate freely without any perturbation, but he
will not see the out-going radiation. For the asymptotic observer, on
the other hand, the Hawking radiation is physically real, and,
due to the non-local interaction between $in$ and $out$-fields,
the in-falling matter will therefore satisfy a non-trivial
time-evolution equation. To make this idea a little bit more concrete,
we will in the last part of this section describe a simple proposal
for a possible effective description of the time-evolution of
the in-falling matter as seen by an asymptotic observer.
The description will be far from complete, but it will give a clear
indication of what kind of modifications from the standard picture
can be expected when one takes this new quantum effect into account.

To simplify the following discussion, let us concentrate on the
time-evolution at late times $\lambda(\T- \T_0)>\!> 1$, in which case
\hleft\ reduces to
\eqn\hlef{H_L(\T) = H - \kappa \lambda^2 (\T-\T_0)}
plus exponentially small corrections. In the $\kappa$-term we recognize
the linear decrease in the left-moving energy as a consequence of
the constant Hawking emission process. The idea is now to define an
effective time evolution in such a way that this expression \hlef\
represents the energy contained in just the left-moving matter.
It turns out that this can be achieved in a very simple and natural way via
the {\it Ansatz} that the hamiltonian $H_L$ not only to {\it measures}
the total remaining energy, but also {\it generates} the effective
time-evolution of the in-falling fields. Thus we define new
left-moving matter fields $f_L(r,t)$ that satisfy the time-evolution
\eqn\evol{-i\partial_t{f}_L(r,t)  = [H_L(t), f_L(r,t)].}
Using \tcomf\ we find that this leads to a modified
free field equation of motion
\eqn\seom{(\partial_t\! -\! \partial_r)f_L(r,t) =
{\kappa\over 2} e^{-\lambda(r+t+\R_0  -\T_0)} (\partial_t\! +\! \partial_r)
f_L(r,t).}
This equation, which is valid for in-falling waves
at super-critical trajectories $r+t >\!> \R_0+\T_0$,
shows that the effective fields $f_L(r,t)$ travel
with a velocity that is slightly bigger than the speed of light!
It can be integrated to a complete
trajectory for the left-moving signals, which takes the
form\footnote{*}{The
integration constant is chosen such that $f_L(r,t) = f_{in}(r+t)$ at
 $r\!-\!t = \R_0\! - \!\T_0$.}
\eqn\solf{f_L(r,t) = f_{in}(
{1\over \lambda} \log[ e^{\lambda (r+t+\R_0  -\T_0)}+
{\kappa \lambda \over 2}
(r\!-\!t - \R_0\! + \!\T_0)]).}
This equation shows that the $f_L$ waves will travel
for a long time along an approximately light-like trajectory,
while they slowly move towards the asymptotic mirror trajectory
$r+t = \R_0+\T_0$.

The special property of the effective time-evolution
is that the total left-moving wave constantly loses energy at
precisely the rate of the Hawking energy flux. A simple
calculation gives that
\eqn\sent{\partial_t\, \Bigl[{1\over 2}
\int ((\partial_t \!+\! \partial_r)f_L)^2 \Bigr] = -
{\kappa \lambda \over 2} \,  \int e^{-\lambda(r+t+\R_0 - \T_0)}
 ((\partial_t \!+\! \partial_r)f_L)^2 = -\kappa \lambda^2}
where the integrals run over $r-t = constant$ and in the second
step we again used \pcs. This demonstrates the consistency of
the interpretation of $H_L$ as the total energy carried by the
effective fields $f_L$.

Another interesting feature of \evol\ is that
the total $x^+$-momentum $P_+$ defined in \pk,
and therefore also the parameter $\T_0$ that determines the asymptotic
mirror trajectory, in fact remain constant.
This tells us that the in-falling fields $f_L$ are capable of catching
up with the mirror in a finite time. When we consider the specific
example of an incoming shock-wave wave of energy $E_{in}$ that
initially falls in along the light-like trajectory
$r_1+t_1>\R_0\!+\!\T_0$, then we find from \solf\ that the wave
will have reached the asymptotic mirror trajectory when
\eqn\limtime{t\! - \! r + \R_0\! - \!\T_0  = {2\over \kappa\lambda}
(e^{\lambda (r_1+t_1+\R_0  -\T_0)} - e^{2\lambda \R_0})
 = {2\over \kappa\lambda^2}(E_{in} - E_c). }
with $E_c$ the critical energy \oc. This is precisely the black
hole evaporation time ${T_{evap} = {1\over \kappa \lambda^2}(E_{in} - E_c)}$
after which the evaporating black hole has reached a sub-critical energy.

Although we have introduced them in a somewhat ad hoc way, all this
suggests to us that these fields $f_L(r,t)$ may indeed
give an reasonable effective description of the infalling
matter as seen by an asymptotic observer, the idea being that
their somewhat unusual time-evolution arises because of the interaction
with the out-going radiation. In any case, the above description
indicates that taking this interaction into account may indeed lead
to a new possible mechanism for stopping the acceleration of the mirror
after all energy has been emitted. It is clear, however, that much work has
to be done to see if an effective description of this kind can be
developed into a complete and consistent semi-classical treatment of
the model at super-critical energies.

\vskip 1cm

\noindent
{\bf Acknowledgements.}

This work is for a large part based on ideas and results that were
developed in collaboration with E. Verlinde. We also acknowledge
discussions with A. Bilal, C. Callan, I. Kogan and K. Schoutens.
This research was financially supported by NSF Grant PHY90-21984.

\listrefs

\end